put
\documentstyle[12pt,a4wide]{article}
\catcode`@=11 
\@addtoreset{equation}{section} 
\def\@eqnnum{\hbox to .01pt{}\rlap{\rm \hskip -\displaywidth\theequation}}
\renewcommand{\theequation}{(\arabic{section}.\arabic{equation})}

\newenvironment{proof}{{\bf Proof:}}{}
\newtheorem{theorem}{Theorem}[section]
\newtheorem{lemma}[theorem]{Lemma}

\newcommand{\qed}{$\bf \Box$}
\newcommand{\Scri}{{\cal I}}
\newcommand{\eps}{\varepsilon}
\newcommand{\be}{\begin{equation}}
\newcommand{\ee}{\end{equation}}
\newcommand{\p}{\partial}
\newcommand{\bea}{\begin{eqnarray}}
\newcommand{\eea}{\end{eqnarray}}

\newcommand{\nn}{\nonumber}
\begin{document}
\title{ The Newtonian Limit on Asymptotically Null Foliations}

\author{Mirta S. Iriondo\thanks{Supported by STINT, The Swedish Foundation for International Cooperation in Research and Higher Education. }
 , Enzo O. Leguizam\'on
\thanks{Fellow of Se.CyT-UNC.}
\and and Oscar A. Reula
\thanks{Member of CONICET.}\\
{\small FaMAF, Medina Allende y Haya de la Torre,}\\
 {\small Ciudad Universitaria, 5000 C\'ordoba, Argentina}\\
{\small Fax nr: 54-51-334054}\\}

\maketitle

\vspace{-.4in}
\begin{abstract}
By suitably re-scaling the conformal Einstein's equations  we are able to apply recent  results in the theory of PDE, and prove that they  possess slow solutions  in a future  neighborhood of an initial surface reaching  ${\cal I}^+$. The structure of the equations obtained 
allows to split (up to any given order) the initial data into those generating slow solutions, i.e., those driven by the sources, and those generating fast solutions, i.e., those which represent gravitational radiation with no relation to the sources. Thus effectively resulting in a proposal to prescribe initial data for solutions with no extra radiation up to the order
needed for each given application.
\end{abstract}

\section{Introduction}

A very important step in the study of  generation of gravitational waves by the selfgravitating motion of celestial bodies or black holes is to determine which initial data must be given in order  to start the computation of the gravitational dynamics. For one would like to specify data with the  minimal amount of gravitational radiation not related to the sources, that is to tailor the initial data for gravity to the data for the sources. This task is particularly difficult in general relativity for the nonlinearities of the theory do not allow for a clear cut resolution into source driven and source independent solutions. 
Fortunately a theory describing in detail the behaviour of solutions to equations with different time scales has been developed in the last two decades which allows to approach the problem in a more systematic way.  For the case of the problem at hand, which 
we have taken to be  an asymptotically flat space-time with matter sources on it, there are two clearly distinct time scales, a slow one which can be associated to the characteristic Newtonian time scales, for instance the orbital period of a test particle circling around the 
sources near their boundaries, that is a distance associated with its mass and size. The most important characteristic of this time scale is that does not change if one increases the value of the speed
of light. The other time scale, faster than the previous, is the time light takes to travel across the source, this clearly changes as the speed of light is increased.
Thus, by changing the value of the speed of light in the equations one is able to distinguish
between these two time scales. 
In turn  these two time scales can be used to distinguish between solutions carrying gravitational 
radiation generated by the sources from those which represent pure radiation, not
coming from any source. The former would move without appreciable change as we
take the speed of light to infinity, while the latter would oscillate faster and faster, thus in the limit becoming singular. But does this difference between solutions exists? and how can we characterize it? Physically we know that something like this must happens for this
is what we impose when we speak of the Newtonian limit, and subsequent postnewtonian approximations, and we know they are a good approximation to describe  some celestial systems. Mathematically the existence and characteristics of this splitting of solutions, which is only
asserted up to any finite given order, is provided by the theory of different time scales mention above, see for example  \cite{kre}, \cite{tad} and \cite{km}.

For these mathematical  theory to apply it it must be assumed that  the system of partial differential equations ruling the dynamics is  symmetric hyperbolic and become singular in a very specific way. If we define the parameter controlling the ratio between the two time scales as $\eps$, in our case the inverse of the speed of light, the systems must be of   the following form
\be
\label{eq:s-h}
A^0{}_{ij}(\eps u^k) \frac{\p}{\p t} u^j =(\frac{1}{\eps}K^a{}_{ij} + A^a_1{}_{ij}(u^k,\eps)
)\partial_a u^j
+ B_{i}(u^k,\eps),
\ee
with the matrices $K^a$  $A^a_1{}$ and $A^0{}$ symmetric,   being $K^a$ constant, $A^0{}$ positive definite and  the matrices $A^a_1{}$ and the vector $B$  continuous in $\eps$ and uniformly continous in $u$ for bounded $u$ (see \cite{rauch1}, \cite{rauch2},\cite{schochet}).

If the system is as above then one can assert that there are one parameter families (in $\eps$) of initial data sets which, for a finite period of time, give raise to solutions whose derivatives up to some given previously specified order are bounded in the limit $\eps \to 0$. Those are the slow solutions one is seeking in order to study the gravitational radiation produced by  sources.
Even more, the theory also tells how the rest of the solutions, called fast solutions  would behave when they are small
(order $\eps$) compared with  slow solutions. The choice of initial data giving slow solutions is called {\it  initialization} and, as the theory shows,  simply consists in making sure that the time derivatives of the solution up to the desired order  are bounded at the initial surface. Using 
the evolution equations these requirements become conditions on the initial data and its spatial derivatives.

In \cite{fri-reula} and \cite{iri-le-reu} the task of writing Einstein's equations
in a form like \ref{eq:s-h} was carried out in the context of a time function associated with an asymptotically flat foliation of space time, thus the assertion that the solutions arising from those special initial data sets had regular limit did not hold all the way up to null
infinity, and so  the results could not be applied to obtain conclusive estimates   about the gravitational radiation produced.

 In order to truly study the radiation phenomena, in this paper we rewrite Einstein's equations in an asymptotically null foliation of space-time, and accomplish the task of casting them in the form \ref{eq:s-h}. This is done in the next section, where we heavily use Friedrich's schema of conformally regularizing Einstein's equations (c.f \cite{helmut}). Once this is done we use,  in the third  section, the resulting system of equations  to obtain the Newtonian limit, which in this context tells how to select initial data sets whose solutions are bounded for a finite time intervall, that is we perform a first order
initialization of the data for the system. It is worth noticing  at this point that besides evolution equations, the  Einstein system has also   constraint equations  between the initial data. They complement and are consistent with the ones the initialization scheme produces, leading together into the Newtonian limit equations. 
\section{The conformal Einstein field equations as a symmetric hyperbolic system}
In order to dimensionalize the Einstein equations  we introduce, as in  \cite{fri-reula}, the concept of dimension for geometrical objects.
We impose that the  coordinate functions   have dimension  $L$
and define the dimension of the components of a vector $n^a$,  $[n^a]$,  such that when the vector acts on any function $f$, it gives a function of dimension $[n(f)]=[n^a]\frac{1}{L}[f]$. We define the dimension of a covector $m_a$,  $[m_a]$,  such that when the covector acts on any vector $n^a$ it gives a function of dimension $[m_an^a]$.  We extend these definitions to tensor fields in the obvious way. This implies that  the metric tensor is dimensionless,  the connection has dimension of $L^{-1}$ and the Riemann tensor has dimension of $L^{-2}$. The  parameter which shall distinguish between slow and fast solutions is, $\eps=c^{-1}$, where $c$ is the velocity of light.

 We shall deal  with a four-dimensional Lorentzian manifold $(\hat M,\hat g_{ab})$\footnote{ Here, we  use the abstract index notation, see \cite{wald}}, where $\hat g_{ab}$ has signature $ (-,+,+,+)$ and satisfies the Einstein's field equations
\be
\hat R_{ab}-\frac{1}{2}\hat R\hat g_{ab}= \kappa{}\eps^4 T_{ab}
\label{einstein}
\ee
with $\kappa{}=8\pi G$,  $G$ been the gravitational constant. Since the dimension of the Ricci tensor is $L^{-2}$,  the energy-momentum tensor $T_{ab}$ has dimensions of energy density.

  Furthermore the source is thought to have  compact support, physically we are thinking in a isolated system, for instance a binary system.  Thus we shall deal with asymptotically flat space-times at future null infinity \cite{geroch}, i.e. a triple $(M,g,\Omega)$ that satisfies:

\begin{enumerate}
\item
$M$ is a four-dimensional manifold with boundary $\Scri^+$, $\Scri^+$ is diffeomorphic to ${\bf R}\times {\bf S}^2$.
\item 
$g_{ab}$ is a Lorentz metric on $M$, $(M,g_{ab})$ is time and space oriented and strongly causal.
\item
$\Omega$ is  a function {\it  (\lq the conformal factor \rq)} on $M$ with
$$
\Omega>0 \quad\mbox{on}\quad \hat M=M\backslash \Scri^+;\qquad \Omega\equiv 0,\quad \mbox{d}\Omega\neq 0 \quad\mbox{on}\quad \Scri^+.
$$
\item
$\Scri^+$ is a null hypersurface with respect to $g_{ab}$ in the past of $\hat M$.
\item
The metric $\hat g_{ab}=\Omega^{-2}g_{ab}$ satisfies the Einstein's field equations \ref{einstein}.
\item
There is a spacelike hypersurface ${\cal S}$ of $(M,g)$ with two-dimensional boundary $Z$ which also belongs to  $\Scri^+$. The hypersurface $S$ is diffeomorphic to the closed unit ball in ${\bf R}^3$ whence $Z$ is diffeomorphic to the sphere \footnote{Most of our resuls are local, so some topological conditions are superfluous.} ${\bf S}^2$.
\end{enumerate}

 Expressing the field equation \ref{einstein} for $\hat g_{ab}$ in terms of $g_{ab}$ and $\Omega$, and splitting them into the trace-free part and the trace yield

\bea
\Omega \sigma_{ab}&=&-\nabla_a\nabla_b \Omega+\frac{1}{4}g_{ab}\nabla^c\nabla_c \Omega+\frac{\kappa{}\eps^4}{2}\;\Omega\stackrel{o}{T}_{ab}
\label{einsteinc1}\\
\Omega^2 R&=&-\kappa{}\eps^4 T -6(\Omega \nabla^c\nabla_c \Omega-2\nabla^c\Omega\nabla_c \Omega),
\label{einsteinc2}
\eea
where $R$ is the Ricci scalar, $2\sigma_{ab}$ the trace-free part of the Ricci tensor and $\nabla$ denotes the Levi-Civita covariant derivative with respect to $g_{ab}$, $\stackrel{o}{T}_{ab}$ is the trace free part of the energy momentum tensor field and $T=\hat g^{ab}T_{ab}$. Remark that since the sources are assumed compact support, $T_{ab}$ has a natural extension to $M$.

Under  the rescaling  $(g,\Omega)\to (\Theta^2 g, \Theta\Omega)$, where the function $\Theta$ is positive everywhere, the Ricci scalars of $g_{\mu\nu}$ and of $\Theta^2g_{\mu\nu}$ are related by
$$
\Theta\;R[g_{\mu\nu}]-\Theta^3\;R[\Theta^2\;g_{\mu\nu}]=6\;\nabla_{\lambda}\nabla^{\lambda}\Theta.
$$
Thus given any Cauchy data for $\Theta$ on an initial surface ${\cal S}$ with $\Theta$ positive on ${\cal S}$, one may determine $\Theta$ in a neighbourhood of ${\cal S}$ such that the Ricci scalar of the rescaled metric becomes constant. We shall use this conformal freedom to set the value of $R$ to be $6\eps^2$.

As a first step to treat the conformal Einstein field equations as a system with two different time scales, we cast them  as a first order symmetric hyperbolic system with constraints, using the formalism developed by Helmut Friedrich in \cite{helmut}.

\subsection{The Choice of Frame}

Following  Friedrich, we choose an  orthonormal frame, this choice yields an  hyperbolic system of  the form  \ref{eq:s-h}, where the matrices $K^{a}$ depend on the frame.
Since we want to describe the Newtonian limit, i.e. the limit when $\eps \to 0$, we need  that the matrices $K^{a}$  be constant in that limit on the hypersurface $\tilde {\cal S}$ on which we give initial data . We achieve this result demanding that the frame  become constant in this limit.

To give a better idea of this  choice of frame we exemplify it in Minkowski space-time. The line element takes the form
$$
\mbox{d}\hat{s}^2=-\frac{\mbox{d}t'^2}{\eps^2}+\mbox{d}r'^2+r'^2\mbox{d}\omega^2,
$$
where $\mbox{d}\omega^2$ denotes the standard line element on the two-dimensional unit sphere. A possible hypersurface on which we could give initial data is the spacelike hyperboloid  given by 
$$
\tilde {\cal S}=\{ t'^2-\eps^2 r'^2=1,\quad t' > 0 \}.
$$

By performing the coordinate transformation
$$
t'+\eps r'= \tan \left( \frac{t+\eps r}{2} \right), \qquad  t'-\eps r'= \tan \left( \frac{t-\eps r}{2} \right),
$$
and rescaling the line element with the conformal factor 
$$
\Omega(t,r)=2  \cos\left( \frac{t+\eps r}{2} \right) \cos\left( \frac{t-\eps r}{2} \right),
$$
one obtains the unphysical line element
\be
\label{Mink}
\mbox{d}s^2=-\frac{\mbox{d}t^2}{\eps^2}+\mbox{d}r^2+\frac{\sin^2 (\eps r)}{\eps^2}\mbox{d}\omega^2.
\ee
Minkowski space-time is only the region
$$
0 < t < \pi,\, 0\leq r,\,  |t+\eps r|\leq \pi,\, |t-\eps r|\leq \pi.
$$

In this new coordinate system the closure of the initial surface $\tilde {\cal S}$ is given by ${\cal S}=\{ t=\pi/2 ,\, 0\leq r \leq \pi/(2\eps) \}$, this surface represents a ball of radius $ \pi/(2\eps)$ and when $\eps \to 0$ it becomes the three-dimensional Euclidean space. Note that the surfaces $t'\pm \eps r'=const.$ represent the advanced and retarded null cones and in the limit  (when $\eps \to 0$) they become the hypersurface $t=const.$,  where $t$ represents now the absolute time.

We claim that we can choose on ${\cal S}$  coordinates such that the  frame  becomes constant when $\eps \to 0$. To see this we note that in Cartesian coordinates  the induced metric on ${\cal S}$ is of the form
$$
\mbox{d}s^2=\mbox{d}x^2+\mbox{d}y^2+\mbox{d}z^2+\eps^2 h_{ij}(\eps,x^i)\mbox{d}x^i\mbox{d}x^j,
$$
for a known smooth in $(\eps, x^i)$ matrix $h_{ij}$. Thus the frame to be choose  is of the form
$$
e_0=\eps\frac{\partial}{\partial t}, \qquad e_i=e^j{}_i\frac{\partial}{\partial x^j},
$$
where $e^i{}_j=\delta^i{}_j+\eps\tilde e^i{}_j$, for some smooth function $\tilde e^i{}_j$. 

Notice that the second fundamental form associated with this surface,  $\chi_{ij}$, vanishes, the conformal factor is $\Omega(\pi/2,r)=2  \cos\left( \frac{\pi/2+\eps r}{2} \right) \cos\left( \frac{\pi/2-\eps r}{2} \right)$ and   defining $\Sigma_0=e_0(\Omega)$ on ${\cal S}$ we have $\Sigma_0=-\eps$. This complete the example.

We now turn into the general case, it is clear that for metrics close to Minkowski we can choose an orthogonal frame $e^a{}_\mu$, with $\mu=0,1,2,3$, in such a way that  the components of the metric in this frame become
$$
g_{\mu\nu}= g_{ab}e^a{}_\mu e^b{}_\nu=\mbox{diag } (-1,1,1,1).
$$
Given a spacelike hypersurface ${\cal S}$, we choose  $e_0$ as the unit normal  to the hypersurface and three orthonormal  vector fields $e_i$ ($i=1,2,3$) on ${\cal S}$, which will be propagated parallel in the direction of $e_0$. Furthermore, we fix Gaussian normal coordinates $x^\alpha$ on a neighborhood of ${\cal S}$ in $D({\cal S})$, hence 

$$
e_0=\eps\frac{\partial}{\partial t}, \qquad e_i=e^\alpha{}_i\frac{\partial}{\partial x^\alpha},
$$
where $e_i$ are tangent to the hypersurfaces $S_{t}=\{t=const.\}$ and $e^j{}_ i|_{\eps=0}=\delta^j{}_i$ 

We denote by $\gamma^\mu{}_{\nu\eta}$ the  connection coefficients  of the canonical covariant derivative $\nabla$ with respect to $e^a{}_\nu$, i.e. 

$$
\nabla_\nu e^a{}_\eta=\gamma^\mu{}_{\nu\eta}e^a{}_\mu 
$$
and with the choice of the frame given above, the connection coefficients have the following properties
$$
\gamma_{0\mu}{}^\nu=0, \qquad\gamma_{\mu(\nu}{}^\eta g_{\lambda)\eta}=0.
$$
 Finally we choose the gauge source as in the Minkowski case, i.e. $R=6\eps^2$. 

\subsection{The Variables}

Following Friedrich's work and after several steps which are detailed in the Appendix, the conformal Einstein equations  can be split into  constraint  and  evolution equations  for the variables 
$$
u=(e^a{}_0,e^a{}_j,\gamma^k{}_{ij}, \chi_{ij},\Omega,\Sigma_0,\Sigma_j,s,\sigma_{ij},\sigma_{0i},E_{ij}, B_{ij}),
$$
where $\Omega E_{ij}$ and $\Omega B_{ij}$ represent the electric and the magnetic parts of  Weyl tensor, $\chi_{ij}=\gamma^0{}_{ij}$ represents the second fundamental form on ${\cal S}$.  The variables $\Sigma_0$,$\Sigma_i$ and $s$, when the conformal Einstein's equations  are satisfied, become $\Sigma_\nu=\nabla_\nu \Omega$ and $ s=\frac{1}{4}\nabla^\mu\nabla_\mu\Omega$ respectively. Note that the variables have been already rescaled in such a way that the source of the system becomes regular in $\eps$ and the Newtonian potential  appears in the first initialization. Thus these variables  correspond to the ``tilde'' variables in the Appendix.

\subsection{The Conformal Evolution Equations:}
\bea
  &&\Omega_{,t}- \Sigma_{0}=0\nn\\
  && s_{,t}-\eps^2  \sigma_{00}  \Sigma_0 +\eps^3 \sigma_{0i} \Sigma_j  g^{ij}+\frac{\kappa{}\eps^2}{8}\left(3\rho  \Sigma_0 -\frac{1}{3}\Omega\rho_{,t}\right)+\nn\\
&&\qquad+\frac{\kappa{}\eps^4}{2}\left(\frac{1}{4} S  \Sigma_0+J^i  \Sigma_i+\frac{1}{12 \Omega}S_{,t}\right)-\frac{1}{4} \Sigma_0=0\nn\\
 &&\Sigma_{0,t}+\eps^2 \Omega  \sigma_{00}+\frac{3}{4}  \Omega+  s -\frac{3\kappa\eps^2}{8}  \Omega^{-1}\rho -\frac{\kappa{}\eps^4}{8}  \Omega^{-1}  S=0\nn\\
 &&\Sigma_{i,t}+\eps^2 \Omega  \sigma_{0i}+\frac{\kappa{}\eps^3}{2}  \Omega J_{i}=0\nn\\
  &&  e^i{}_{l,t}+  \chi^i{}_l+\eps  \chi^j{}_l\,  e^i{}_j=0\nn\\
 && e^a{}_{0,t}=0\label{eq:evol2}\\
&& \chi_{i j, t}+\eps \chi_{l j} \chi^l{}_i -\eps \Omega   E_{ij}-\eps\left( \sigma_{ij}-  g_{ij}  \sigma_{00}\right)=0\nn\\
&& \gamma^i{}_{j k, t}+\eps^2 \gamma^i{}_{l k} \chi^l{}_{j}-\eps  \Omega  B_{j\tau}\epsilon^{\tau i}{}_k+\eps\left(\delta^i{}_{j} \sigma_{0k}-  g_{kj}  \sigma^i{}_{0}\right)=0\nn\\
 && \sigma_{0 i,t}-\frac{1}{\eps}\nabla^{j} \sigma_{ij}=0\nn\\
 && \sigma_{i j,t}-\frac{1}{\eps}\nabla_{(i} \sigma_{j)0}- \Sigma_0   E_{ij}-\eps  \Sigma_l \;  B_{\tau (i}\;\epsilon^{\tau l}{}_{j)}-\frac{1}{\eps}  \Omega \kappa  t_{0(ij)}=0\nn\\
 &&  E_{ij,t} +\frac{1}{\eps} D_k   B_{l(i} \epsilon_{j)}{}^{kl}-3\eps  E^k{}_{(i}   \chi_{j)k}  +2\eps  \chi_k{}^k   E_{ij}+\eps  g_{ij}   E_{km}   \chi^{km}+\frac{1}{\eps} \kappa  t_{0(ij)}=0\nn\\
 && B_{ij,t} -\frac{1}{\eps} D_k   E_{l(i} \epsilon_{j)}{}^{kl}-3\eps   B^k{}_{(i}   \chi_{j)k} + 2\eps  \chi_k{}^k   B_{ij}+\eps  g_{ij}   B_{km}   \chi^{km} +\frac{\kappa}{2\eps} \epsilon^{nl}{}_{(i}  t_{j)nl} =0\nn
\eea
where $D_k$ is the covariant derivative on the hypersurfaces ${\cal S}$ and  $\rho$, $J^i$,  $\eps^2 S^{ij}$ and $t_{\mu\nu\lambda}$, represent, respectively,  the energy density, the momentum density, the pressure tensor  and derivatives of the energy momentum tensor, as seen by an observer at rest with respect to the foliation chosen. They are defined by 
\be
\label{tab}
T^{ab}=\rho \;e^a{}_0e^b{}_0+2\eps\;J^{(a}e^{b)}{}_0+\eps^2\;S^{ab},
\ee
and 
\be
\label{deft}
t_{\lambda\eta\nu} :=\Omega^{-1}\hat \nabla_{[\lambda}F_{\eta]\nu}=\Omega^{-1}\left(\nabla_{[\lambda} F_{\eta]\nu}+\Omega^{-1}\Sigma_{[\lambda} F_{\eta]\nu}-\Omega^{-1}g_{\nu[\lambda} F_{\eta]\beta}\Sigma^\beta\right),
\ee
being $F_{\eta\nu}=\frac{1}{2}\left(T_{\eta\nu}-\frac{1}{3}\hat g_{\eta\nu}T\right)$. In spite of the presence of $\frac{1}{\eps}$  in the source terms of the three last equations of \ref{eq:evol2} it can be seen that they are smooth in $\eps$ (see Appendix).

This system is of the form \ref{eq:s-h}, as can be seen after a lengthy and mostly uninteresting calculation. Thus if appropriately coupled with symmetric hyperbolic equations describing the matter fields\footnote{Here we are assuming two things, first that the matter equations are not singular in $\eps$ and second that boundary conditions for the matter fields can be handled appropriately in order for the usual energy estimates to hold, (the case of electromagnetism  can be treated without further problems by adding it to the singular sector).} it results in a bigger system with the same properties, see for example \cite{helmut1}. Thus the general theory of systems with different time scales can be applied in a straight forward way.

\subsection{The Conformal Constraint Equations:}

The projection of the  conformal Einstein equations on ${\cal S}$ yields the following constraint equations:
\bea
&&D_i \Omega-\eps^2 \Sigma_i=0\nn\\
&&D_i   s -\eps^3 \sigma_{i0} \Sigma_0+\eps^4 \sigma_{ij} \Sigma^j-\frac{\kappa{}\eps^4}{2}J_{i} \Sigma_0-\frac{\kappa{}\eps^6}{2}\stackrel{o}{S}_{ij} \Sigma^j-\frac{\kappa\eps^2}{24 \Omega}D_i \rho+\frac{\kappa{}\eps^4}{24 \Omega}D_i S +\frac{3\eps^2}{4} \Sigma_i=0\nn\\
&&D_i \Sigma_0-\eps^3 \chi_{ij} \Sigma^j+\eps^3  \Omega \sigma_{i0}+\frac{\kappa{}\eps^4}{2} \Omega J_{i}=0\nn\\
&&D_i \Sigma_j-\eps \chi_{ij} \Sigma_0+\eps^2 \Omega \sigma_{ij}+\frac{ \Omega}{4}  g_{ij}-  s  g_{ij}-\frac{\kappa{}\eps^4}{2} \Omega\stackrel{o}{S}_{ij}=0\nn\\
&&2  e^i_{[k,j]}+\eps(  e^i_{k,l}  e^l_{j}-  e^i_{j,l}  e^l_{k})-\eps( \gamma^i{}_{jk}- \gamma^i{}_{kj})=0
\label{constraint2}\\
&&{}^3  r^i{}_{jkl}+2\eps^2 \chi^i{}_{[k}  \chi_{l]j}-2\eps^2 \Omega \{\delta^i{}_{[k}  E_{l]j}-l_{j[k}  E_{l]}{}^i\} -2\eps^2\delta^i{}_{[k}  \sigma_{l]j}-2\eps^2
  g_{j[l}  \sigma^i{}_{k]}-\delta^i{}_{[k}  g_{l]j}=0\nn\\
&&2D_{[j}  \chi_{k]i}-\eps^2 \Omega \;  B_{i\tau}\;\epsilon^\tau{}_{jk}-2 \eps^2  g_{i[j}  \sigma_{k]0}=0\nn\\
&&2D_{[i} \sigma_{j]0}-2\eps^2 \chi^k{}_{[i} \sigma_{j]k}-\eps^2 \Sigma_k \;  B^k{}_\tau \;\epsilon^\tau{}_{ij}- \Omega \kappa  t_{ij0}=0\nn\\
&&2D_{[i} \sigma_{j]k}-2\eps^2 \chi_{k[i} \sigma_{j]0}-2\eps^2 \Sigma_l \{\delta^l{}_{[i}  E_{j]k}-l_{k[i}  E_{j]}{}^l\}-\eps \Sigma_0\;  B_{k\tau}\;\epsilon^\tau{}_{ij}-  \Omega\kappa t_{ijk} =0\nn\\
&&D^i  E_{ij}-\eps^2  B_{ik}\; \chi^i{}_l\;\epsilon^{kl}{}_{j}+\kappa \;t_{0j0}=0\nn\\
&&D^i   B_{ij}+\eps^2  E_{ik}\; \chi^i{}_l\;\epsilon^{kl}{}_{j}+\frac{\kappa}{2}\;t_{mn0}\epsilon^{mn}{}_j=0,\nn
\eea
where ${}^3  r^i{}_{j k l}$  is the Riemann tensor of the hypersurface $t=const.$ given by
$$
{}^3 r^i{}_{j k l}=e_k ( \gamma^i{}_{l j})-e_l( \gamma^i{}_{k j})+\eps^2\left( \gamma^i{}_{k m} \gamma^m{}_{l j}- \gamma^i{}_{l m} \gamma^m{}_{k j}- \gamma^i{}_{m j}( \gamma^m{}_{k l}- \gamma^m{}_{l k})\right
)
$$
and $\stackrel{o}{S}_{ij}$ is the traceless part of $S_{ij}$.  We shall assume for  the rest of the paper that there exist smooth in $\eps$ one parameter families of solutions to the constraint equation that are $C^p$, for some $p>1$. The existence and regularity of the solutions to the constraint equations have been studied  in \cite{ACF}, and the smoothness of the solutions in the parameter $\eps$ should follow from the implicit function theorem.

\section{The Newtonian limit}

In the present context the Newtonian limit is defined as the subset of the one parameter family (in $\eps$) of solutions to  the equation system \ref{eq:evol2}-\ref{constraint2} which are bounded as $\eps$ goes to zero. 

 According to the theory of different time scales, to achieve this result it is enough to initialize the equation system, i.e. to require that  the time derivative of the dynamical variables on ${\cal S}$ remains finite when $\eps$ goes to zero, being this requirement consistent with the constraint equations. Since the solutions will remain bounded in time, as the theory of different time scales asserts, it implies that these solutions  will satisfy similar initialization and constraint equations  on each future hypersurface of the foliation.

Assuming  the dynamical variables are $C^1$ in $\eps$, and that the initial data can be set as
$$
u(x,0)=u_0(x)+{\cal O}(\eps),
$$
the singular part of the evolution equations \ref{eq:evol2} with the above requirement  yields at the initial surface \footnote{Since these equations will be valid for each slice ${\cal S}_t$, $t\in [t_0,T]$.} ${\cal S}$ 
\bea
\label{eq:smooth}
&&D^{j} \sigma_{ij}=0\nn\\
&& D_{i} \sigma_{0 j}=0\nn\\
&& D_{k}   B_{l(i}{}\epsilon^{kl}{}_{j)}=0\\
&& D_{k}   E_{l(i}{}\epsilon^{kl}{}_{j)}=0.\nn
\eea
and to zeroth order the constraint equations \ref{constraint2} become

\bea
&&D_i \Omega= 0 \nn\\
&&D_i {}   s=0 \nn\\
&&D_i \Sigma_j=  (s-\frac{\Omega}{4})   g_{ij}\nn\\
&&D_i \Sigma_0= 0\nn\\
&& e_{[k,j]} = 0 \label{constraint3}\\
&&{}^3  r^i{}_{jkl} =\delta^i{}_{[k}g_{l]j}\qquad \mbox{(Gauss' equation)}\nn\\
&&D_{[j}  \chi_{k]i} = 0 \qquad\mbox{(Codacci's equation)}\nn\\
&&D_{[i} \sigma_{j]0} =  0\nn\\
&&D_{[i}(\sigma_{j]k} - g_{j]k}\frac{k}{12}\rho)= 0 \nn\\
&&D^i(  E_{ij}-\frac{\kappa}{6 \Omega}   g_{ij} \rho) = 0 \nn\\
&&D^i   B_{ij} = 0. \nn
\eea

Notice that the $\eps=0$ limit of the geometry at which we solve the initialization equations is rather singular. If we picture the compactification as immersed in the Einstein Universe\footnote{This is just a convenient picture and is not totally correct, for in the presence of matter we know that at least the point corresponding to $i^0$ must be singular}, then the part of the Cauchy surface of the Einstein Universe corresponding to our initial surface, becomes bigger and bigger, and in fact each Cauchy surface of Einstein Universe with topology $S^3$ in the limit blows up to become ${\bf R}^3$.

We stress that the general relativistic dynamics, which is
what we are interested in,  after all, has a non-vanishing $\eps$. Thus, we shall be interested in the solutions to the initialization equations in just a portion of ${\bf R}^3$, that part that corresponds with our initial surface.

If no boundary conditions are imposed on the solutions to the initialization equations an infinite set of solutions to them would be possible and so it would result in a non unique Newtonian limit. We believe this undeterminancy is due to the fact that the initial surface is not a Cauchy surface and that most of these solutions would in fact  not arise from  solutions which are asymptotically flat at space-like infinity,  for they would not have the correct (mild)  singular structure there. We conjucture that the only solutions arising from an asymptotically flat space at space-like infinity are those which when considered as solutions in the whole limiting initial slice (${\bf R}^3$ with a flat metric) have asymptotically flat boundary conditions and,  in what follows we shall treat just that case.

We solve now the  system \ref{eq:smooth}-\ref{constraint3}. The conformal freedom in the initial data is fixed as in Minkowski space-time, namely
\be
\label{eq:init}
  \Omega=1,\quad     s=-\frac{3}{4},\quad\mbox{and}\quad \Sigma_i=-x_i.
\ee
When we fixed the value of the Ricci scalar we still had a freedom left in the choice of the time derivative of $\Theta$ at the initial surface ${\cal S}$ and since  the mean  curvature of ${\cal S}$ changes as 
$$
\chi=\Theta^{-1}(\hat \chi+3\;\frac{\partial\Theta}{\partial x_0}),
$$
under a conformal transformation, we choose $\displaystyle{\frac{\partial\Theta}{\partial x_0}}$,  and in turn $\Sigma_0$,  in such a way that the rescaled mean curvature becomes zero. 

Consider now the equations for $\chi_{ij}$, fixing $j$, these equations say that $\chi_{ij}=D_iV_j$. Because of the symmetry of $\chi_{ij}$ we have $D_{[i}V_{j]}=0$ which in turns implies that $V_i=D_i\;\psi$, for some arbitrary function $\psi$. Thus if $\chi=0$, then
$$
\Delta \psi=0,
$$
imposing  the asymptotic condition that the field vanishes at infinity, $\psi$ must be zero.

Since  $ D_{i} \sigma_{0 j}=0$,  imposing the same asymptotic condition as above,  we must have $\sigma_{0j}=0$.

Finally, under appropiate boundary conditions we claim that the unique solutions  of the equations for  $B_{ij}$, $  \sigma_{ij}$ and $E_{ij}$ in \ref{eq:smooth} and \ref{constraint3} are given by 
$$
 B_{ij}=0,\quad \sigma_{ij}=-\frac{1}{3}D_iD_j\phi_\sigma+g_{ij}\frac{\kappa}{12}\rho \quad 
\mbox{and}\quad E_{ij}=D_i D_j \phi_E -\frac{1}{3} g_{ij}\;                                                                                                                 \Delta \phi_E
$$
where the functions $\phi_\sigma$ and $ \phi_E$  satisfy
$$
\Delta \phi_\sigma=2\pi G \rho.
$$
and 
$$
\Delta \phi_E=2\pi G \rho,
$$
respectively.
Uniqueness of the solution in the Laplacian equation yields $\phi=\phi_E=\phi_\sigma$. We call  this function the Newtonian Potential, and the initial data we have determined the Newtonian data.

It remains to prove the uniqueness claimed above for  the initialization, hence of the Newtonian limit. We do that by showing that the solution for $\sigma_{ij}, E_{ij}$ and $B_{ij}$ are unique. 

It is easily seen that the  difference between two different solutions for the equation for $\sigma_{ij}$ is a symmetric tensor that satisfies $ D^i \delta\sigma_{i j}=0$ and $D_{[i}\delta\sigma_{j]k}=0$. As for $\chi_{ij}$, the symmetry of $\sigma_{ij}$ and the last equation imply that $\delta\sigma _{ij}=D_iD_j \psi$. The first equation implies that 
$$
\Delta D_i\psi=0,
$$
imposing  the asymptotic conditions we have that $\psi=const.$ and so that $\delta\sigma_{ij}=0$. The uniqueness for 
$E_{ij}$ and $B_{ij}$ follows from the next lemma as applied to the difference between two solutions

\begin{lemma} Let $V_{ij}$ be a smooth tensor field in ${\bf R}^3$ such that $V_{ij} \to 0$ quickly enough when $r \to \infty$. Moreover let $V_{ij}$ be a symmetric, transverse and  traceless tensor satisfying
\be
\label{potential}
D_{k}   V_{l(i}{}\epsilon^{kl}{}_{j)}=0.
\ee
Then $V_{ij}=0$. \end{lemma}
\begin{proof}
Using de Rham's Theorem we note that if $V_{ij}$ is a transverse tensor (i.e its divergence is zero) then there exists a tensor field $W_{nm}$ such that
\be
\label{rote}
V_{ij}=D_{k}   W_{li}{}\epsilon^{kl}{}_{j}.
\ee
Since $V_{ij}$ is symmetric, \ref{rote}  implies
\be
\label{divergence}
D_l\;W^{nl}-D^n\; W=0,
\ee
where $W=g^{ij}\;W_{ij}$. Inserting  \ref{rote} in \ref{potential}  we obtain
\be
\label{eq:ws}
\Delta\; W_{(ij)}-D^m\;D_{(i}\; W_{|m|j)}=0.
\ee
 Since  $V_{ij}$ and $W_{ij}$ satisfy linear equations  it is sufficient to prove that $V_{ij}(W^S)=V_{ij}(W^A)=0$, where $W^S$ ($W^A$)  is the symmetric (antisymmetric) part of $W$.

 We consider first $V_{ij}(W^A)$, equation \ref{rote} implies that 
$$
D_l\; k^l=0
$$
where $k^l=\epsilon^{lmn}\;W_{mn}=\epsilon^{lmn}\;W^A{}_{mn}$. Then
$$
V_{ij}(W^A)=-D_i\; k_j.
$$
The symmetry of  $V_{ij}$  implies that $k_j=D_j \psi$ for some scalar function $\psi$, and since  $D_l\; k^l=0$, $\psi$ satisfies  $\Delta \psi=0$. Thus $\psi=const.$ , which in turns implies  $V_{ij}(W^A)=0$.

Consider now the symmetric part $W^S$. Using equation \ref{divergence}, equation \ref{eq:ws} becomes
\be
\label{eq:w}
\Delta\; W^S_{ij}-D_i\;D_j\; W^S=0.
\ee
Let $\phi$ such that $\Delta\phi=W^S$, where $W$ is the trace of $W^S_{ij}$, then \ref{eq:w} becomes  
$$
\Delta(W^S_{ij}-D_iD_j \phi)=0,
$$
using the asymptotic conditions, uniqueness of solutions to Laplace equation implies,
$$
W^S_{ij}=D_iD_j \phi.
$$
Inserting this expression for $W^S$ into \ref{rote}, we obtain $V_{ij}(W^S)=0$. \qed \end{proof}

\section{Conclusions}
By suitably rescaling Friedrich version of Einstein equations we have cast them in a form where standard results of the theory of different time scales apply. In particular this theory asserts that there exist slow solutions, that is solutions which move at the pace of the matter motion and so describe the gravitational radiation they produce. Even more the theory tells  how to single out these solutions by imposing conditions the initial data sets must satisfy. These conditions are usually called initialization conditions and they are naturally consistent with the constraint equations of General Relativity.

Imposing these conditions to first order we find what in the present setting should be called the Newtonian limit: all degrees of freedom are frozen, only remains a scalar function, the Newtonian Potential, which is completely determined by the matter content of the space-time.

 These initialized solutions remain slow all the way into a finite neighborhood of $\Scri^+$, thus the radiation they imprint at $\Scri^+$ can be truly assigned to the slow motion of the matter sources.

The structure of the equations used allows to control the slowness of the solution to any order on $\eps$, by continuing the initialization procedure i.e. making sure that further
time derivatives of the solution at the initial hypersurface remain bounded. These postnewtonian orders should give enough information as to make a rigorous justification of the Quadrupolar formula, in a similar scheme to the one used by Winicour on null hypersurfaces \cite{Winicour}. But we believe that the most important application to
the above system of equations should be in numerical calculations for there, besides giving a symmetric hyperbolic set of equations, a result already know from Friedrich's work, one would have a tight control on the initial data to be prescribed.

\section{Appendix: The dimensionalized  field equations}
\label{calculations}

In order to  have the  Conformal Einstein field equations as a symmetric hyperbolic system with smooth sources in $\eps$ we shall rescale the variables used in \cite{helmut}.
These variables are 
$$
u=(e^a{}_\mu,\gamma^\mu{}_{\nu\eta}, \Omega,\Sigma_\nu,s,\sigma_{\mu\nu},d^\mu{}_{\nu\eta\lambda}),
$$
where $\Omega d^\mu{}_{\nu\eta\lambda}$ is the Weyl tensor. In these variables the equations \ref{einsteinc1} and \ref{einsteinc2} becomes

\be
z=(O_\mu,P_\mu,Q_{\mu\nu},T^\mu{}_{\nu\lambda},K^\mu{}_{\nu\lambda\eta},L_{\nu\lambda\eta},H_{\nu\lambda\eta})=0,
\label{eqfield}
\ee
where
\bea
O_\mu&=&\nabla_\mu\Omega-\Sigma_\mu\nn\\
P_\mu&=&\nabla_\mu s +\sigma_{\mu\nu}g^{\nu\eta}\Sigma_\eta +\frac{\Omega}{24}\nabla_\mu R+\frac{R}{12}\Sigma_\mu+\nn\\
&&-\frac{\kappa{}\eps^4}{2}\stackrel{0}{T}_{\mu\nu}g^{\nu\eta}\Sigma_\eta+\frac{\kappa{}\eps^4}{24\Omega}\nabla_\mu T\nn\\
Q_{\mu\nu}&=&\nabla_\mu\Sigma_\nu+\Omega\sigma_{\mu\nu}-sg_{\mu\nu}-\frac{\kappa{}\eps^4}{2}\Omega\stackrel{0}{T}_{\mu\nu}\label{eqz}\\
\qquad\quad T^\mu{}_{\nu\lambda}\, e^a_\mu&=&(\gamma^\mu{}_{\nu\lambda}-\gamma^\mu{}_{\lambda\nu})\,e^a_\mu-(e^a_{\lambda,\beta}e^\beta_{\nu}-e^a_{\nu,\beta}e^\beta_{\lambda})
\nn\\
K^\mu{}_{\nu\lambda\eta}&=&r^\mu{}_{\nu\lambda\eta}-R^\mu{}_{\nu\lambda\eta}\nn\\
L_{\nu\lambda\eta}&=&2\nabla_{[\lambda}\sigma_{\eta]\nu}-
\Sigma_\mu d^\mu{}_{\nu\lambda\eta}-\Omega \kappa{}\eps^4 t_{\lambda\eta\nu}+\frac{1}{12}\nabla_{[\lambda}Rg_{\eta]\nu}\nn\\
H_{\nu\lambda\eta}&=&\nabla_\mu d^\mu{}_{\nu\lambda\eta}-\kappa{}\eps^4 t_{\lambda\eta\nu}\nn
\eea
where $R^\mu{}_{\nu\lambda\eta}$  denotes  the Riemann tensor defined by  the metric, and $r^\mu{}_{\nu\lambda\eta}$  the Riemann tensor defined by the connection and the torsion $T^\mu{}_{\nu\lambda}$, i.e

$$
r^\mu{}_{\nu\lambda\eta}=e_\lambda (\gamma^\mu{}_{\eta\nu})-e_\eta(\gamma^\mu{}_{\lambda\nu})+\gamma^\mu{}_{\lambda\beta}\gamma^\beta{}_{\eta\nu}-\gamma^\mu{}_{\eta\beta}\gamma^\beta{}_{\lambda\nu}-\gamma^\mu{}_{\beta\nu}(\gamma^\beta{}_{\lambda\eta}-\gamma^\beta{}_{\eta\lambda}-T^\beta{}_{\lambda\eta})
$$

Besides the unknown $u$, the function $R$ appears in the equation system $z=0$. This is a gauge source function for the conformal factor which shall be chosen in what follows to be constant as in the Minkowski case, i.e.  $R=6\eps^2$.

$Q_{\mu\nu}=0$ entails that the trace free part of the conformal Einstein's equation \ref{einsteinc1} is satisfied, $T^\mu{}_{\nu\lambda}=0 $ and $K^\mu{}_{\nu\lambda\eta}=0$ guarantee that the metric reconstructed from the frame is the canonical one. 
$L_{\nu\lambda\eta}=0$ and $H_{\nu\lambda\eta}=0$ are the Bianchi identities with matter and  $P_\mu=0$  is an integrability condition for $Q_{\mu\nu}=0$, i.e. it is obtained by taking the covariant derivative of $Q_{\mu\nu}$, contracting and using $L_{\lambda\eta\nu}=0$(see \cite{helmut1}). 

Instead of the variables $d^\mu{}_{\nu\lambda\eta}$ we shall use the variables $E_{\tau\sigma}$ and $B_{\tau\sigma}$. These tensor fields can be thought as  the electric and magnetic parts  of the Weyl tensor, i.e.

\be
\Omega^{-1}C_{\mu\nu\lambda\rho}=2\{l_{\mu[\lambda}{}E_{\rho]\nu}-l_{\nu[\lambda}{}E_{\rho]\mu}-n_{[\lambda}{}B_{\rho]\tau}{}\eps^\tau{}_{\mu\nu}-n_{[\mu}{}B_{\nu]\tau}{}\eps^\tau{}_{\lambda\rho}\}
\ee
where 
\bea
E_{\tau\sigma}&=&\Omega^{-1}C_{\mu\nu\lambda\rho}{}h^\mu{}_{\tau}\;n^\nu\;h^\lambda{}_{\sigma}\;n^\rho\nn\\
B_{\tau\sigma}&=&\Omega^{-1}C^*_{\mu\nu\lambda\rho}{}h^\mu{}_{\tau}\;n^\nu\;h^\lambda{}_{\sigma}\;n^\rho\nn\\
l_{\mu\nu}&=&h_{\mu\nu}+n_{\mu}{}n_{\nu}\nn\\
h_{\mu\nu}&=&g_{\mu\nu}+n_{\mu}{}n_{\nu}\nn\\
\epsilon_{\tau\sigma\eta}&=&\epsilon_{\mu\nu\lambda\rho}n^\mu\nn\\
\Omega^{-1}C^*_{\mu\nu\lambda\rho}&=&\frac{1}{2}\Omega^{-1}C_{\mu\nu\alpha\beta}\;\epsilon^{\alpha\beta}{}_{\lambda\rho}\nn
\eea
here we shall take  $n_\mu$ to be  the unit one form normal to the hypersurfaces $t=const.$, i.e. $n^a=e_0{}^a$.

 Equation \ref{eqfield} can easily be split into the propagation equations along $e_0$ and into the constraint equations projected on $S$. 
\vspace{.2in}

\noindent{\bf The Evolution Conformal Equations:}
By the gauge assumptions, $\nabla_0=\eps \partial_t$, so 

\bea
0&=&O_0=\eps\Omega_{,t}-\Sigma_{0}\nn\\
0&=&P_0=\eps s_{,t}+\sigma_{0\nu}\Sigma_\mu g^{\mu\nu}-\frac{\kappa{}\eps^4}{2}\stackrel{o}{T}_{0\nu}\Sigma_\mu g^{\mu\nu}+\frac{\kappa{}\eps^5}{24\Omega} T_{,t}+\frac{\eps^2}{2}\Sigma_0\nn\\
0&=&Q_{0\nu}=\eps\Sigma_{\nu,t}+\Omega\sigma_{0\nu}-sg_{0\nu}-\frac{\kappa{}\eps^4}{2}\Omega\stackrel{0}{T}_{0\nu}\nn\\
0&=&T^\mu{}_{0 \lambda}\, e^a_\mu= \eps e^a_{\lambda,t}+\gamma^\mu{}_{\lambda 0}\,e^a_\mu\nn\\
0&=&K^\mu{}_{\nu 0 \beta}=\eps\gamma^\mu{}_{\beta \nu, t}+\gamma^\mu{}_{\eta \nu}\gamma^\eta{}_{\beta 0}-\Omega \left(-e_0{}^\mu E_{\beta\nu} +e_{0\nu}E_{\beta}{}^\mu+B_{\beta\tau}\epsilon^{\tau\mu}{}_\nu\right)-\label{eq:evol}\\
&&-2\delta^\mu{}_{[0}\sigma_{\beta]\nu}-2g_{\nu[\beta}\sigma^\mu{}_{0]}-\eps^2\delta^\mu{}_{[0}\;g_{\beta]\nu}\nn\\
0&=&-g^{\nu \lambda}L_{\nu\lambda i}=\eps\sigma_{0 i,t}-\nabla^{j}\sigma_{ij}+\Omega \kappa{}\eps^4 g^{\nu \lambda}t_{\lambda i \nu}\nn\\
0&=&L_{i 0 j}=\eps\sigma_{i j,t}-\nabla_{i}\sigma_{0 j}-\Sigma_0 E_{ij}-\Sigma_lB_{i\tau}\epsilon^{\tau l}{}_j-\Omega \kappa{}\eps^4 t_{0ij}\nn\\
0&=&H_{(i|0|j)}=\eps  E_{ij,t} + D_k B_{l(i} \epsilon_{j)}{}^{kl}-3E^k{}_{(i} \chi_{j)k} + 2\chi_k{}^kE_{ij}+\nn\\
&&+g_{ij} E_{km}\chi^{km} + \kappa \eps^4 t_{0(ij)}\nn\\
0&=&-\frac{1}{2}H_{nl(i}\epsilon^{nl}{}_{j)}=\eps B_{ij,t} - D_k E_{l(i} \epsilon_{j)}{}^{kl}-3B^k{}_{(i} \chi_{j)k}  + 2\chi_k{}^kB_{ij}+\nn\\
&&+g_{ij} B_{km}\chi^{km}  + \frac{\kappa}{2} \eps^4 \epsilon^{nl}{}_{(i}t_{j)nl}.\nn
\eea

\noindent{\bf The Conformal Constraint Equations:}

\bea
0&=&O_i=D_i\Omega-\Sigma_i\nn\\
0&=&P_i=D_i s +\sigma_{i\mu}\Sigma^\mu -\frac{\kappa{}\eps^4}{2}\stackrel{0}{T}_{i\nu}\Sigma^\nu+\frac{\kappa{}\eps^4}{24\Omega}D_i T+\frac{\eps^2}{2}\Sigma_i \nn\\
0&=&Q_{i0}=D_i\Sigma_0-\chi_{ij}\Sigma^j+\Omega\sigma_{i0}-\frac{\kappa{}\eps^4}{2}\Omega\stackrel{0}{T}_{i0}\nn\\
0&=&Q_{ij}=D_i\Sigma_j-\chi_{ij}\Sigma_0+\Omega\sigma_{ij}-sg_{ij}-\frac{\kappa{}\eps^4}{2}\Omega\stackrel{0}{T}_{ij}\nn\\
0&=&T^i{}_{jk}=(\gamma^i{}_{jk}-\gamma^i{}_{kj})-(e^i_{k,l}e^l_{j}-e^i_{j,l}e^l_{k})\nn\\
0&=&K^i{}_{jkl}={}^3r^i{}_{jkl}+2\chi^i{}_{[k}\chi_{l]j}-2\Omega \{\delta^i{}_{[k}E_{l]j}-l_{j[k}E_{l]}{}^i\}\label{constraint}\\
&&\qquad \; -2\delta^i{}_{[k}\sigma_{l]j}-2g_{j[l}\sigma^i{}_{k]}-\eps^2\delta^i{}_{[k}\;g_{l]j}\qquad \mbox{(Gauss' equation)}\nn\\
0&=&K^0{}_{ijk}=2D_{[j}\chi_{k]i}-\Omega B_{i\tau}\epsilon^\tau{}_{jk}-2g_{i[j}\sigma_{k]0}\nn\\
&&\mbox{(Codacci's equation)}\nn\\
0&=&L_{0ij}=2D_{[i}\sigma_{j]0}-2\chi^k{}_{[i}\sigma_{j]k}-\Sigma_k B^k{}_\tau \epsilon^\tau{}_{ij}-\Omega \kappa{}\eps^4 t_{ij0}\nn\\
0&=&L_{kij}=2D_{[i}\sigma_{j]k}-2\chi_{k[i}\sigma_{j]0}-2\Sigma_\mu \{h^\mu{}_{[i}E_{j]k}-l_{k[i}E_{j]}{}^\mu\}+\nn\\
&&\qquad\; -\Sigma_0\;B_{k\tau}\epsilon^\tau{}_{ij}-\Omega \kappa{}\eps^4 t_{ijk}\nn\\
0&=&H_{0mn}\epsilon^{mn}{}_i=D^i E_{ij}-B_{ik}\chi^i{}_l\epsilon^{kl}{}_{j}+\kappa \eps^4 t_{0j0}\nn\\
0&=&H_{0j0}=D^i B_{ij}+E_{ik}\chi^i{}_l\epsilon^{kl}{}_{j}+\frac{\kappa\eps^4}{2}\epsilon^{mn}{}_j t_{mn0},\nn
\eea
\vspace{.2in}

Since the variables $B_{ij}$ and $E_{ij}$ in this gauge (c.f. \cite{Helmut2}) satisfy a symmetric hyperbolic system of differential equations, the system above is, as in the case with the variables $d^i{}_{jkl}$, a symmetric hyperbolic system with constraints.

The sources of this system are not smooth in $\eps$, therefore 
in order to apply the existence theorems,  we  rescale the variables as follows.
\be
\label{resc}
\begin{array}{ccccccc}
& \tilde s=\displaystyle{\frac{s}{\eps^2}},\; &\tilde \Omega=\Omega,\quad \tilde g_{\mu\nu}=g_{\mu\nu},\quad
&\tilde \Sigma_0=\displaystyle{\frac{\Sigma_0}{\eps}}, \quad
&\tilde \Sigma_i=\displaystyle{\frac{\Sigma_i}{\eps^2}},\quad &\tilde \gamma^i{}_{j k}=\displaystyle{\frac{\gamma^i{}_{j k}}{\eps^2}}\\
&e^i{}_j=\delta^i{}_j+\eps\tilde e^i{}_j, &\tilde E_{ij} =\displaystyle{\frac{E_{ij}}{\eps^4}},\quad &\tilde B_{ij}=\displaystyle{\frac{B_{ij}}{\eps^4}}, 
&\tilde \sigma_{\mu\nu}=\displaystyle{\frac{\sigma_{\mu\nu}}{\eps^2}}, &\tilde \chi_{ij}=\displaystyle{\frac{\chi_{ij}}{\eps^2}},\\
&  \eps^2 \;\tilde \sigma_{ij}=-\displaystyle{\frac{\tilde g_{ij}}{4}}+\displaystyle{\frac{\sigma_{ij}}{\eps^2}}. & & & &
\end{array}
\ee
Part of this rescaling is based on the behaviour of some of these variables in  Minkowski space-time  with  metric given by \ref{Mink}. With it  the Conformal Evolution equations become

\bea
0&=&\tilde \Omega_{,t}-\tilde\Sigma_{0}\nn\\
0&=&\tilde s_{,t}-\eps^2\tilde \sigma_{00}\tilde \Sigma_0 +\eps^3\tilde\sigma_{0i}\tilde\Sigma_j\tilde g^{ij}+\frac{\kappa{}\eps^2}{8}\left(3\rho\tilde \Sigma_0 -\frac{1}{3\tilde\Omega}\rho_{,t}\right)+\nn\\
&&+\frac{\kappa{}\eps^4}{2}\left(\frac{1}{4} S\tilde \Sigma_0+J^i\tilde \Sigma_i+\frac{1}{12\tilde\Omega}S_{,t}\right)-\frac{1}{4}\tilde\Sigma_0\nn\\
0&=&\tilde\Sigma_{0,t}+\eps^2\tilde\Omega \tilde\sigma_{00}+\frac{3}{4}\tilde \Omega+\tilde s -\frac{3\kappa\eps^2}{8}\tilde \Omega^{-1}\rho -\frac{\kappa{}\eps^4}{8}\tilde \Omega^{-1}  S\nn\\
0&=&\tilde\Sigma_{i,t}+\eps^2\tilde\Omega \tilde\sigma_{0i}+\frac{\kappa{}\eps^3}{2}\tilde \Omega J_{i}\nn\\
0&=&  \tilde e^i{}_{l,t}+\tilde \chi^i{}_l+\eps\tilde \chi^j{}_l\,\tilde e^i{}_j\nn\\
0&=&  e^a{}_{0,t}\label{eq:evol1}\\
0&=&\tilde\chi_{i j, t}+\eps\tilde\chi_{l j}\tilde\chi^l{}_i -\eps\tilde\Omega \tilde E_{ij}-\eps\left(\tilde\sigma_{ij}-\tilde g_{ij}\tilde \sigma_{00}\right)\nn\\
0&=&\tilde\gamma^i{}_{j k, t}+\eps^2\tilde\gamma^i{}_{l k}\tilde\chi^l{}_{j}-\eps\tilde \Omega\tilde B_{j\tau}\epsilon^{\tau i}{}_k+\eps\left(\delta^i{}_{j}\tilde\sigma_{0k}-\tilde g_{kj}\tilde \sigma^i{}_{0}\right)\nn\\
0&=& \tilde\sigma_{0 i,t}-\frac{1}{\eps}\nabla^{j}\tilde\sigma_{ij}+\frac{1}{\eps}\tilde\Omega \kappa{} g^{\nu \lambda}t_{\lambda i \nu}\nn\\
0&=&\tilde \sigma_{i j,t}-\frac{1}{\eps}\nabla_{(i}\tilde\sigma_{j)0}-\tilde\Sigma_0 \tilde E_{ij}-\eps\tilde \Sigma_l \;\tilde B_{\tau (i}\;\epsilon^{\tau l}{}_{j)}-\frac{1}{\eps}\tilde \Omega \kappa  t_{0(ij)}\nn\\
0&=& \tilde E_{ij,t} +\frac{1}{\eps} D_k \tilde B_{l(i} \epsilon_{j)}{}^{kl}-3\eps\tilde E^k{}_{(i} \tilde \chi_{j)k}  +2\eps\tilde \chi_k{}^k \tilde E_{ij}+\eps\tilde g_{ij} \tilde E_{km} \tilde \chi^{km}+\frac{1}{\eps} \kappa  t_{0(ij)}\nn\\
0&=&\tilde B_{ij,t} -\frac{1}{\eps} D_k \tilde E_{l(i} \epsilon_{j)}{}^{kl}-3\eps \tilde B^k{}_{(i} \tilde \chi_{j)k} + 2\eps\tilde \chi_k{}^k \tilde B_{ij}+\eps\tilde g_{ij} \tilde B_{km} \tilde \chi^{km} +\frac{\kappa}{2\eps} \epsilon^{nl}{}_{(i}  t_{j)nl} \nn
\eea
and the Conformal Constraint equations, 

\bea
0&=&D_i\tilde\Omega-\eps^2\tilde\Sigma_i\nn\\
0&=&D_i \tilde s -\eps^3\tilde\sigma_{i0}\tilde\Sigma_0+\eps^4\tilde\sigma_{ij}\tilde\Sigma^j-\frac{\kappa{}\eps^4}{2}J_{i}\tilde\Sigma_0-\frac{\kappa{}\eps^6}{2}\stackrel{o}{S}_{ij}\tilde\Sigma^j-\frac{\kappa\eps^2}{24\tilde\Omega}D_i \rho+\frac{\kappa{}\eps^4}{24\tilde\Omega}D_i S +\frac{3\eps^2}{4}\tilde\Sigma_i\nn\\
0&=&D_i\tilde\Sigma_0-\eps^3\tilde\chi_{ij}\tilde\Sigma^j+\eps^3 \tilde\Omega\tilde\sigma_{i0}+\frac{\kappa{}\eps^4}{2}\tilde\Omega J_{i}\nn\\
0&=&D_i\tilde\Sigma_j-\eps\tilde\chi_{ij}\tilde\Sigma_0+\eps^2\tilde\Omega\tilde\sigma_{ij}+\frac{\tilde\Omega}{4}\tilde g_{ij}-\tilde s\tilde g_{ij}-\frac{\kappa{}\eps^4}{2}\tilde\Omega\stackrel{o}{S}_{ij}\nn\\
0&=&2\tilde e^i_{[k,j]}+\eps(\tilde e^i_{k,l}\tilde e^l_{j}-\tilde e^i_{j,l}\tilde e^l_{k})-\eps(\tilde\gamma^i{}_{jk}-\tilde\gamma^i{}_{kj})
\label{constraint1}\\
0&=&{}^3\tilde r^i{}_{jkl}+2\eps^2\tilde\chi^i{}_{[k}\tilde \chi_{l]j}-2\eps^2\tilde\Omega \{\delta^i{}_{[k}\tilde E_{l]j}-l_{j[k}\tilde E_{l]}{}^i\} -2\eps^2\delta^i{}_{[k}\tilde \sigma_{l]j}-2\eps^2
\tilde g_{j[l}\tilde \sigma^i{}_{k]}-\delta^i{}_{[k}\tilde g_{l]j}\nn\\
0&=&2D_{[j}\tilde \chi_{k]i}-\eps^2\tilde\Omega \;\tilde B_{i\tau}\;\epsilon^\tau{}_{jk}-2 \eps^2\tilde g_{i[j}\tilde \sigma_{k]0}\nn\\
0&=&2D_{[i}\tilde\sigma_{j]0}-2\eps^2\tilde\chi^k{}_{[i}\tilde\sigma_{j]k}-\eps^2\tilde\Sigma_k \;\tilde B^k{}_\tau \;\epsilon^\tau{}_{ij}-\tilde\Omega \kappa  t_{ij0}\nn\\
0&=&2D_{[i}\tilde\sigma_{j]k}-2\eps^2\tilde\chi_{k[i}\tilde\sigma_{j]0}-2\eps^2\tilde\Sigma_l \{\delta^l{}_{[i}\tilde E_{j]k}-l_{k[i}\tilde E_{j]}{}^l\}-\eps\tilde\Sigma_0\;\tilde B_{k\tau}\;\epsilon^\tau{}_{ij}-\tilde \Omega\kappa t_{ijk} \nn\\
0&=&D^i\tilde E_{ij}-\eps^2\tilde B_{ik}\;\tilde\chi^i{}_l\;\epsilon^{kl}{}_{j}+\kappa \;t_{0j0}\nn\\
0&=&D^i \tilde B_{ij}+\eps^2\tilde E_{ik}\;\tilde\chi^i{}_l\;\epsilon^{kl}{}_{j}+\frac{\kappa}{2}\;t_{mn0}\epsilon^{mn}{}_j,\nn
\eea

We claim that in the system \ref{eq:evol1} the source is smooth in $\eps$. This  is not apparent  at a first sight because  the four last equations have singular sources. But using energy-momentum conservation, we see that  the projections of the nonphysical divergence  (i.e. the conformal continuity equations) become :
\bea
\rho_{,t}+D^i{}J_i&=&\tilde \Omega^{-1}\tilde\Sigma_0\rho-\eps \tilde\chi\;\rho-\eps^3 S^l{}_n\;\tilde\chi^n{}_l+\tilde \Omega^{-1}(2\eps^2\tilde\Sigma_i{}J^i+\eps^2\tilde\Sigma_0\;S)\\
J_{i,t}+D^l{}S_{il}&=&\tilde\Omega^{-1}\rho\tilde\Sigma_i+2\tilde\Sigma_0\;J_i+\tilde\Omega^{-1}\eps^2\;(2\tilde\Sigma^j{}S_{ij}-S\tilde\Sigma_i)-\eps\tilde\chi\;J_i+\eps\tilde\chi^l{}_i\;J_l.\nn
\eea

Consequently, inserting \ref{tab} in  definition \ref{deft} for $t_{\mu\nu\lambda}$ and using the continuity equations,  we obtain:
\bea
2\tilde \Omega t_{0j0}&=&-\frac{1}{3} D_j\rho-\frac{2}{3}\eps^2\rho\tilde\Sigma_j+\frac{\eps^2}{2}\left(D^iS_{ij}-\frac{1}{3}\tilde g_{ij}S\right)-\eps^2\tilde\Omega^{-1}\tilde\Sigma_0\;J_j+\\
&&-\frac{\eps^3}{2}\tilde\chi_j{}^l\;J_l-\frac{\eps^4}{2}\tilde\Omega^{-1}\left(\tilde\Sigma^lS_{lj}-\frac{1}{3}S\tilde\Sigma_j\right)+\frac{\eps^3}{2}\tilde\chi\;J_j\nn\\
4\tilde \Omega t_{0(ij)}&=&\eps D_{(i}J_{j)}+\frac{\eps}{3}\tilde g_{ij}\left(\rho_{,t}-\tilde\Omega^{-1}\rho\tilde\Sigma_0\right)+\eps^2\rho\tilde\chi_{ij}+\\
&&+\eps^3\left(S_{ij,t}-\frac{1}{3}\tilde g_{ij}S_{,t}\right)+\eps^3\tilde\Omega^{-1}\tilde\Sigma_0\left(S_{ij}-\frac{1}{3}\tilde g_{ij}S\right)+\nn\\
&&+\eps^3\tilde\Omega^{-1}\left(\tilde\Sigma_{(i}J_{j)}-\tilde g_{ij} J_l\tilde\Sigma^l\right)-\frac{\eps^3}{3}\tilde\Omega^{-1}\tilde g_{ij} S\tilde\Sigma_0+\eps^4\tilde\chi_{(i}{}^lS_{j)l}\nn\\
2\tilde \Omega t_{ij0}&=&\eps D_{[j}J_{i]}-\eps^2\tilde\Omega^{-1}\tilde\Sigma_{[i}J_{j]}-\eps^4\tilde\chi_{[i}{}^lS_{j]l}\\
2\tilde \Omega t_{ijk}&=&\frac{1}{3}D_{[i}\rho\; \tilde g_{j]k}+\eps^2\tilde\Omega^{-1}\tilde\Sigma_{[i}\left(\frac{1}{3}\rho\tilde g_{j]k}+\eps^2(S_{j]k}-\frac{1}{3}S\tilde g_{j]k})\right)+\\
&&+\eps^2\; D_{[i}\left( S_{j]k}-\frac{1}{3}S\tilde g_{j]k}\right)-\tilde\Omega^{-1}\tilde\Sigma_0\eps^2\tilde g_{k[i}J_{j]}-\nn\\
&&-\eps^2\tilde\Omega^{-1}\tilde\Sigma^l\tilde g_{k[i}\left(\frac{1}{3}\rho\tilde g_{j]l}+\eps^2(S_{j]l}-\frac{1}{3}S\tilde g_{j]l})\right)+\eps^3\tilde\chi_{k[i}J_{j]}.\nn
\eea

Using these expressions it is easy to see that
$$
\tilde g^{\nu\mu}t_{\nu i \mu}=0
$$
and 
$$
\epsilon^{nl}{}_{(i} t_{j)nl} ={\cal O}(\eps^2),
$$
showing that the apparently singular terms are in fact regular.

\end{document}